# Determinants of local chemical environments and magnetic moments of high-entropy alloys

Bo Li, Wang Gao,* and Qing Jiang

Key Laboratory of Automobile Materials, Ministry of Education, Department of Materials Science and Engineering, Jilin University 130022, Changchun, China.

High-entropy alloys (HEAs) such as CrMnFeCoNi exhibit unconventional mechanical properties due to their compositional disorder. However, it remains a formidable challenge to estimate the local chemical-environment and magnetic effects of HEAs. Herein we identify the state-associated cohesive energy and band filling originated from the tight-binding and Friedel models as descriptors to quantify the site-to-site chemical bonding and magnetic moments of HEAs. We find that the s-state cohesive energy is indispensable in determining the bonding-strength trend of CrMnFeCoNi that differs from the bonding characteristics of precious and refractory HEAs, while the s-band filling is effective in determining the magnetic moments. This unusual behavior stems from the unique chemical and magnetic nature of Cr atoms and is essentially due to the localized and transferred itinerant electrons. Our study establishes a fundamental physical picture of chemical bonding and magnetic interactions of HEAs and provides a rational guidance for designing advanced structural alloys.

High-entropy alloys (HEAs), as a novel class of metallic materials, can be divided into different categories in terms of their components, such as precious HEAs (e.g. IrPdPtRhRu), refractory HEAs (e.g. NbMoTaTiV), 3d transition HEAs (e.g. CrMnFeCoNi) and so on. These alloys often possess highly desirable combinations of mechanical properties that are promising for many practical applications[1–6], because of their disordered multicomponent structures. CrMnFeCoNi HEAs, one of the most promising HEAs to date, possess the unique site-to-site chemical and magnetic disorder and exhibit superior damage tolerance and fracture toughness[7–11]. However, the nearly randomly distributed atomic species of HEAs limit understanding the origin of their unique mechanical behavior. Therefore, a fundamental issue of HEAs in general and CrMnFeCoNi in particular is how to estimate quantitatively the local chemical-environment and magnetic effects.

Several important attempts have been made to address this issue. Atomic-scale short-range order (SRO) parameters are widely used in studying the chemical and magnetic distribution of HEAs[12]. However, they are mainly suitable for describing the statistical properties instead of the site-to-site chemical bonding and magnetic moments. The magnetic frustration of antiferromagnetic Cr atoms is expected to reduce the Cr-Cr pairs in the nearest-neighbor (1nn) shell of CrCoNi medium-entropy alloys and CrMnFeCoNi HEAs[13–18]. On the other hand, a Cr atom has the biggest covalent force among the components of CrMnFeCoNi HEAs, indicating that the Cr-Cr pairs generally show the strongest chemical bonding and should promote the structural stability. There thus seems to exist a competition between the magnetic frustration and chemical bonding of the Cr-Cr pairs for determining the structure of CrMnFeCoNi HEAs. Furthermore, the conventional Slater-Pauling curve demonstrates that the average magnetic moments are linearly correlated with the valence-electron number, which is mainly applicable to transition metals (TMs) and their bi-metallics such as FeV, FeCo, and FeCr alloys[19–21]. However, CrMnFeCoNi HEAs have been recommended as a candidate to break the Slater-Pauling curve[22]. In addition, the valence-electron concentration, electronegativity difference, atomic size difference, entropy of mixing, and enthalpy of mixing are also proposed as descriptors to study the phase stability and deformation mechanism of HEAs[2,23–26]. However, these empirical parameters mainly focus on the average bulk-related properties but are inconvenient to capture the local chemical-environment and magnetic effects.

By using the state-associated cohesive properties and band filling based on the tight-binding (TB) and Friedel models, we have proposed the mixed d- and s-state cohesive energy and the s-band filling as novel descriptors for characterizing the local chemical bonding and magnetic moments of CrMnFeCoNi HEAs. We find that CrMnFeCoNi HEAs present extraordinary bonding characteristics compared to precious and refractory HEAs and exhibit unconventional Cr-dominant local magnetic moments, which stem from the unique role of Cr atoms. Notably, it is the localized and transferred itinerant electrons that change the d-states of Cr atoms and connect the magnetic moments between Cr atoms and other atoms. Our scheme uncovers the fundamental origin of chemical and magnetic interactions of HEAs, which is helpful for designing advanced structural alloys.

## Results
**The descriptor construction of chemical environments and magnetic moments of HEAs.** The TB and Friedel models[27,28] based on one-electron approximation provide an effective solution to quantitatively estimate the contribution of different bands to chemical bonding and magnetic moments of TMs. The s- and d-state components of cohesive energies ($E_{coh\text{-}s}$ and $E_{coh\text{-}d}$) and the s- and d-band filling ($n_s$ and $n_d$) are obtained analytically by the density of s- and d-type states[27,28] (see more details in Supplementary Note 1). It demonstrates that the d-electrons control the variation of chemical bonding from one TM to the next, while the free-like s-electrons make a minor contribution. The Slater-Pauling curve[19,20] demonstrates that the total band filling ($n$) is effective in capturing the average magnetic moments of bi-metallics and the d-bands govern the magnetic properties of bi-metallics. It should be mentioned that the value of $n_s$ based on the Friedel model[28] is different from that based on the conventional band theory.

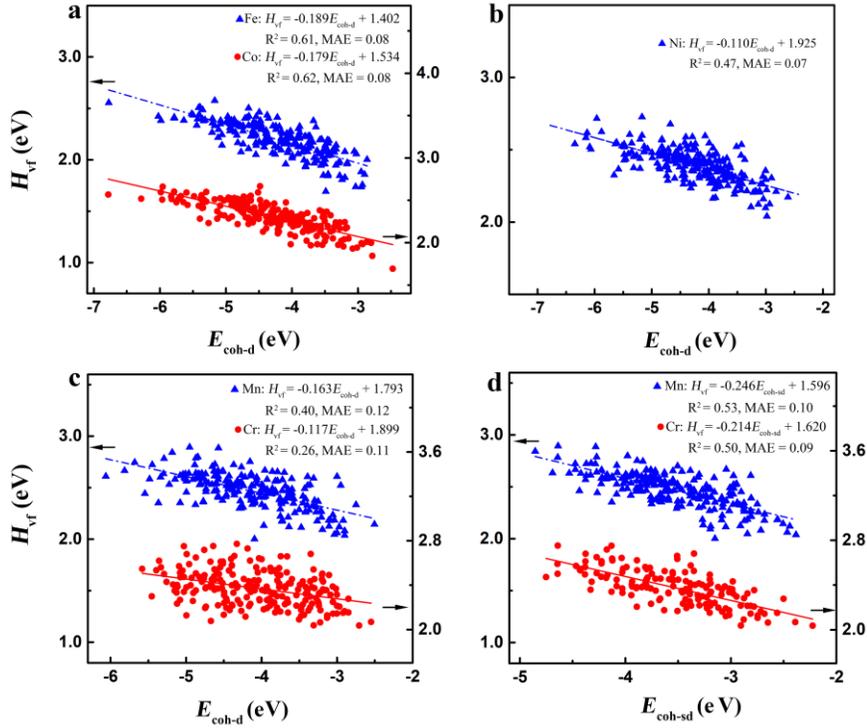

**Figure 1. The non-spin-polarized $H_{vf}$ as a function of the state-associated cohesive energies in CrMnFeCoNi HEAs. a** and **b,** The Fe, Co and Ni's $H_{vf}$ against the d-state cohesive energy ($E_{coh-d}$). **c** and **d,** Comparison between the $E_{coh-d}$ and $E_{coh-sd}$ (involving the s-state cohesive energy of Cr atoms and the d-state cohesive energy of other atoms) in describing Cr and Mn's $H_{vf}$.

The former for 3d TMs varies from 0.46 for Sc to 0.84 for Fe (Supplementary Table 1), while the latter for 3d TMs is about 0.6.

We thus employ the state-associated cohesive properties and band filling to characterize the chemical bonding and magnetic effects of HEAs, which are the cohesive energy $E_{coh}$ and band filling $n$ and their s- and d-state components ($E_{coh-s}$, $E_{coh-d}$, $n_s$, and $n_d$) (Supplementary Note 1). For a given site in HEAs, its complex chemical-environment and magnetic effects are quantified with the geometric mean of cohesive energy and band filling of its surrounding atoms, as,

$$E_{coh} = \left(\prod_i^N E_{coh,i}\right)^{1/N} \quad (1)$$

$$n = \left(\prod_i^N n_i\right)^{1/N} \quad (2)$$

where $E_{coh,i}$ and $n_i$ are the cohesive energy and band filling of $i$th involved atom in the bulk state and $N$ is the number of the involved atoms. Similarly, the corresponding s- and d-state contribution as well as the mixed s- and d-state contribution can be also obtained. Accordingly, one can define the descriptors for the 1nn shell, second-nearest-neighbor (2nn) shell, third-nearest-neighbor (3nn) shell, fourth-nearest-neighbor (4nn) shell, and the rest long-range shell for a given site in HEAs.

**The determinants of chemical environments of HEAs.** We now study the effect of chemical environments of HEAs, by first focusing on the non-spin-polarized vacancy formation energy ($H_{vf}$) of CrMnFeCoNi HEAs. Fig. 1 plots the correlation between $H_{vf}$ and the cohesive-energy descriptors (obtained with atoms in the vacancies' 1nn shell). Apparently, $H_{vf}$ shows a linear relation with the d-state cohesive energy $E_{coh-d}$ for Fe, Co, and Ni vacancies with reasonable accuracy (Fig. 1a and b), where the error percentage is only about 10%. In contrast, the s-state cohesive energy $E_{coh-s}$ fails to capture the trends of $H_{vf}$ (Supplementary Fig. 1a and b). Hence, it is the d-states that dominate the chemical-environment effect of Fe, Co, and Ni sites in CrMnFeCoNi HEAs. However, both $E_{coh-d}$ and $E_{coh-s}$ perform poorly in describing the $H_{vf}$ of Cr and Mn vacancies (Fig. 1c and Supplementary Fig. 1c). Interestingly, $E_{coh-sd}$ (involving the s-state cohesive energy of 1nn Cr atoms and the d-state cohesive energy of other 1nn atoms) describes well the $H_{vf}$ of Cr and Mn sites (Fig. 1d). This seems to contradict the conventional TB and Friedel models[27,28] that the d-states control the cohesive properties of TMs. In particular, for the interatomic bonding, the Cr atoms' behavior depends strongly on their surrounding atoms in CrMnFeCoNi HEAs: the Cr atoms bond to Fe, Co, and Ni atoms mainly with the d-states and likely bond to Cr and Mn atoms mainly with the s-states. The underlying mechanism will be discussed in detail below.

Furthermore, the $H_{vf}$ of different vacant species can be correlated by combining the removed vacant atoms' cohesive energies ($E_{coh,\ rva}$) and their environment descriptors ($E_{coh,\ 1nn}$), as $D_{CE} = k_1 \times E_{coh,\ rva} + k_2 \times E_{coh,\ 1nn}$, where $k_1$ and $k_2$ reflect the coupling between the removed vacant atom and its nearest neighbors. Fig. 2a shows the relation of $H_{vf}$ against $D_{CE}$: except Ni sites, Fe and Co sites' $H_{vf}$ comply with a consistently linear relation, whereas Cr and Mn sites do another one. This implies the similar bonding characteristics for Fe and Co atoms and for Cr and Mn atoms in CrMnFeCoNi HEAs. We also study the $H_{vf}$ of IrPdPtRhRu, NbMoTaTiV, and NbMoTaTiW HEAs[29]. In

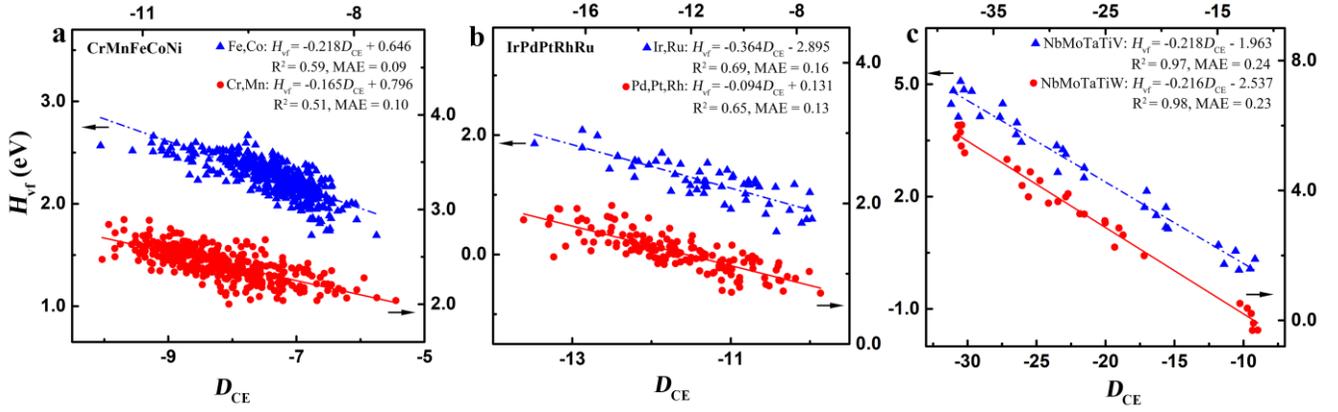

**Figure 2.** The non-spin-polarized $H_{vf}$ as a function of the combined descriptor ($D_{CE}$) of the removed vacant atom and its nearest neighbors. **a,** Fe and Co vacancies and Cr and Mn vacancies of CrMnFeCoNi HEAs. **b,** Ir and Ru vacancies and Pd, Pt, and Rh vacancies of IrPdPtRhRu HEAs. **c,** All vacancies of NbMoTaTiV and NbMoTaTiW HEAs[29].

Fig. 2b and c, these HEAs' $H_{vf}$ can also be well described by $D_{CE}$ that obtained by only using the d-state cohesive energies of the corresponding components, in stark contrast with the Cr and Mn vacancies of CrMnFeCoNi HEAs. Furthermore, the five elements of IrPdPtRhRu HEAs are divided into two groups (Ir and Ru and Pd, Pt and Rh in Fig. 2b), whereas those of NbMoTaTiV or NbMoTaTiW HEAs show an identical behavior (Fig. 2c). This reveals another significant difference between the considered HEAs that the bonding type between the different components is in the order of CrMnFeCoNi HEAs (3 types) > IrPdPtRhRu HEAs (2 types) > NbMoTaTiV and NbMoTaTiW HEAs (1 type).

Note that we also attempt to use the electronic descriptor $\psi = \left(\prod_{i=1}^{N} S_{v_i}\right)^{2/N} / \left(\prod_{i=1}^{N} \chi_i\right)^{1/N}$ ($S_{v_i}$ and $\chi_i$ represent the valence-electron number and electronegativity of $i$th atom in a given atom's 1nn shell) to describe the non-spin-polarized $H_{vf}$ of HEAs, which reflects the d-band upper edge[30–33]. We find that $\psi$ performs better than $E_{coh-d}$ in determining the $H_{vf}$ of IrPdPtRhRu HEAs but does worse in that of CrMnFeCoNi HEAs (Fig. 1 and Supplementary Figs. 2 and 3). Hence, the chemical-environment effect of CrMnFeCoNi HEAs is more controlled by the d-band width (the cohesive energy) while that of IrPdPtRhRu HEAs is governed by the d-band upper edge.

These findings clarify several crucial controversies and offer basic insights for the chemical bonding of HEAs. Fig. 1 shows that $H_{vf}$ of CrMnFeCoNi HEAs linearly decreases with increasing of the cohesive-energy descriptor, implying that the sites with more small-cohesive-energy (the large absolute value) nearest neighbors should have enhanced stability. In particular, the component of the cohesive-energy descriptors is in the order of $E_{coh-d, Mn} < E_{coh-d, Fe} < E_{coh-s, Cr} < E_{coh-d, Co} < E_{coh-d, Ni}$ for the Cr sites, different from that of the bulk TM d-state cohesive energy. Note that the Cr-Cr pairs are found to present magnetically repulsive interactions[13–18]. Hence, Cr-Cr pairs are detrimental to the stability of CrMnFeCoNi HEAs from the view of both chemical bonding and magnetic frustration, eliminating the generally considered contradiction that chemical bonding and magnetic frustration of Cr-Cr pairs compete with each other. On the other hand, the sites with more large-cohesive-energy (the small absolute value) nearest neighbors like Ni are more prone to form vacancies in CrMnFeCoNi HEAs (Fig. 1), in agreement with the findings by Huang et al and Chen et al that vacancies in CrCoNi, CrCoFeNi, and CrMnFeCoNi alloys are more likely to generate at the sites with a Ni-rich 1nn shell[34,35]. Moreover, our scheme also suggests that the sites with more 1nn-Pd atoms of IrPdPtRhRu HEAs and more 1nn-Ti atoms of NbMoTaTiV and NbMoTaTiW HEAs have the low stability. This rationalizes the experimental results[29] that Ti atoms in NbMoTaTiV HEAs are thermodynamically unstable and form a dendritic structure and in NbMoTaTiW HEAs occasionally lead to a dual-phase bcc+hcp structure. Overall, our findings demonstrate that the nearest neighbors of a given site significantly affect its stability, supporting the existence of SRO. Notably, the SRO parameters are generally adopted to understand the chemical distribution of HEAs due to their statistical nature[12–18], whereas our scheme effectively quantifies the chemical bonding of HEAs by considering the bonding strength between a given atom and its nearest neighbors.

**The determinants of magnetic moments of HEAs.** We now turn to understand the magnetism of CrMnFeCoNi HEAs. To elucidate the local magnetic effect, a 108-atom supercell is divided into 27 different tetrahedrons, which are the minimum repeat unit of an fcc crystal (Supplementary Fig. 4). Fig. 3, Supplementary Fig. 5 and Supplementary Table 2 show the correlation between the average magnetic moments of each tetrahedron ($M_{tetra}$) of a given bulk structure and the corresponding band filling $n_s$ and $n$. Clearly, the s-band filling $n_s$ performs much better than the total band filling $n$ in describing the local magnetic moments of CrMnFeCoNi HEAs. We also consider the average magnetic moments of each tetrahedron at the same position of different bulk structures. The good linear relation between $M_{tetra}$ and $n_s$ further demonstrates the effectiveness of $n_s$ (Supplementary Fig. 6 and Supplementary Table 3). This is particularly encouraging, as the complex magnetic ordering of CrMnFeCoNi HEAs can be quantitatively captured by a simple electronic descriptor.

Furthermore, we also study the average magnetic moments of the atoms around a vacancy site in CrMnFeCoNi HEAs. Remarkably, the local magnetic moments around vacancies (vacancies' 1nn atoms, $M_{1nn}$) are still well-characterized by $n_s$ rather than $n$ (Fig. 4). In addition, the average magnetic moments of vacancies' 2nn shell ($M_{2nn}$), 3nn shell ($M_{3nn}$), 4nn shell ($M_{4nn}$), and long-range shell ($M_{long}$) are also linearly related to the corresponding $n_s$ (Supplementary Figs. 7 and 8). Notably, the Slater-Pauling curve demonstrates that the average magnetic moment of TMs and bi-metallics is dominated by the d-band filling[21,22], which seems to be at odds with our results in CrMnFeCoNi HEAs. By analyzing the values of $n_s$, one can find that Fe, Co, Ni, and Mn have the similar $n_s$, which is significantly different from that of Cr (Supplementary Table 1). Therefore, it is the Cr atoms that dominate the variation of local magnetic moments of CrMnFeCoNi HEAs. Indeed, the local magnetic-moment trends can be well characterized by only using the number of Cr atoms ($N_{Cr}$) for CrMnFeCoNi HEAs with or without vacancies (Fig. 5). It is noteworthy that the dominant effect of Cr atoms is not due to the significant change of Cr atoms' magnetic moments. $N_{Cr}$ still dictates the magnetic-moment trends even with the less change of Cr atoms' magnetic moments (Supplementary Fig. 9). Ritchie et al[18] proposed that the magnetic moments of CrCoNi ternary alloys are mainly related to the SRO of Cr atoms or Cr-Cr pairs. Hence, the characteristics of Cr atoms control the magnetic moments of both CrCoNi and CrMnFeCoNi alloys.

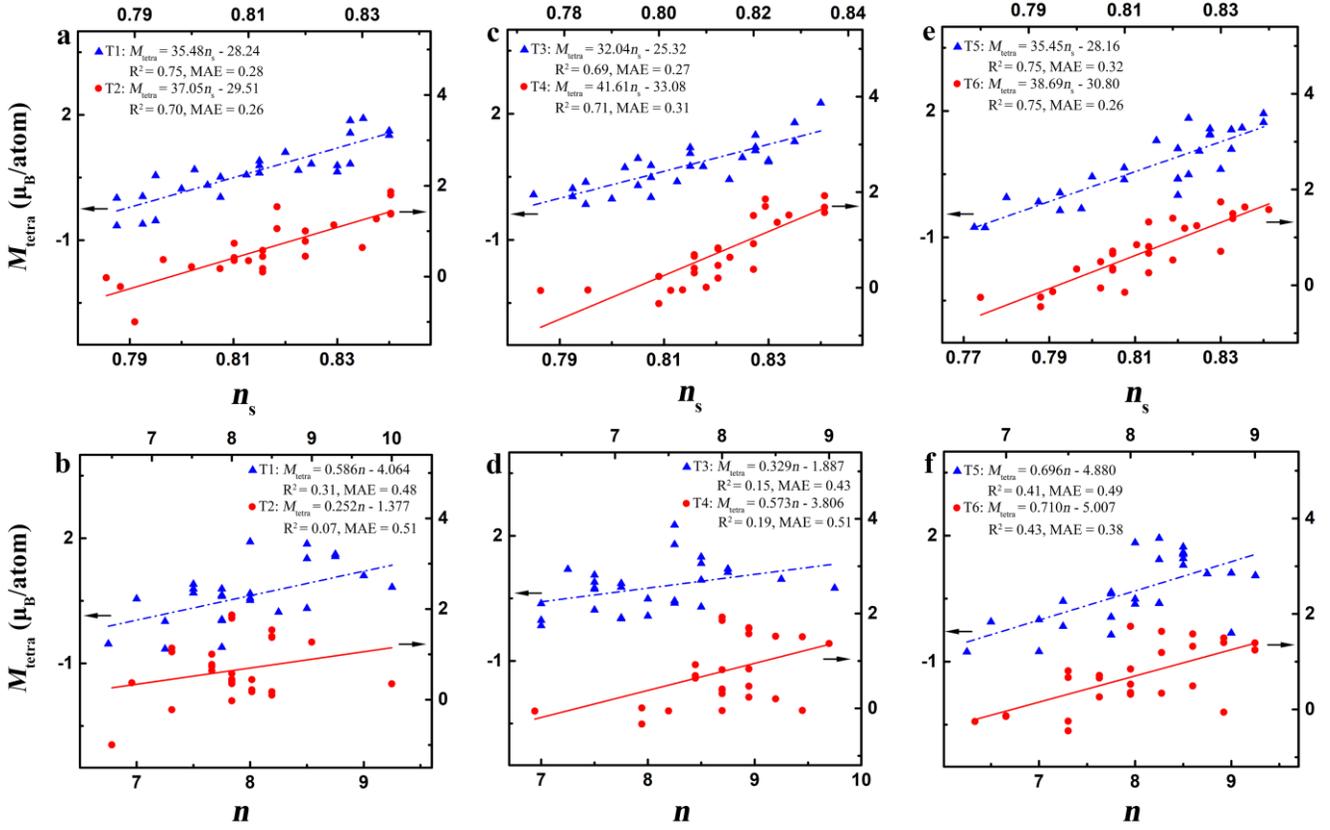

**Figure 3.** Comparison between the s-band filling ($n_s$) and total band filling ($n$) in describing the local magnetic moments of CrMnFeCoNi HEAs. **a-f**, Tetrahedrons of six different bulk structures T1-T6.

Since the chemical environments and magnetic moments together determine the stability of CrMnFeCoNi HEAs, we also attempt to characterize the spin-polarized $H_{vf}$ of CrMnFeCoNi HEAs by using the linear combination of the environment-effect descriptor $E_{coh,1nn}$ and magnetic moments ($M_{1nn}$, $M_{2nn}$, $M_{3nn}$, $M_{4nn}$, $M_{long}$, the vacant atoms' magnetic moments $M_{site}$ and the total average magnetic moments $M_{all}$). Supplementary Fig. 10 and Supplementary Table 4 show that the magnetic moments exhibit the different range-reaching properties in determining the spin-polarized $H_{vf}$, depending on the surrounding chemical distributions of vacancies. Taking Cr vacancies as an example, the sites of the bulk-3 structure that have more Ni and Co nearest neighbors exhibit the largest values of $E_{coh-sd}$, corresponding to a localized magnetic effect. Thus the linear combination of $E_{coh-sd}$ and $M_{1nn}$ performs well in describing the $H_{vf}$ of Cr vacancies (Supplementary Fig. 10c). Similarly, the sites of the bulk-1 structure possess the smallest values of $E_{coh-sd}$, and thus the linear combination of $E_{coh-sd}$ and long-range magnetic moments $M_{long}$ or $M_{all}$ describe well the corresponding $H_{vf}$ (Supplementary Fig. 10a). More details are shown in Supplementary Note 2. Overall, the chemical distributions determine the role of magnetic effects in the stability of CrMnFeCoNi HEAs, supporting the existence of SRO. Thus, both the chemical bonding and magnetic effects indicate the SRO in HEAs.

**Discussion**

Our findings uncover a novel physical picture of the chemical-environment and magnetic nature of HEAs, which can be understood in the framework of the itinerant electron model[36]. The itinerant electrons, different from the localized d-electrons and free-electron-like s-electrons, refer to the d-electrons located around the Fermi level and are widely used to explain the magnetism of TMs and their bi-metallics. The electron localization function (ELF) shows that there are electrons localized mainly between Cr

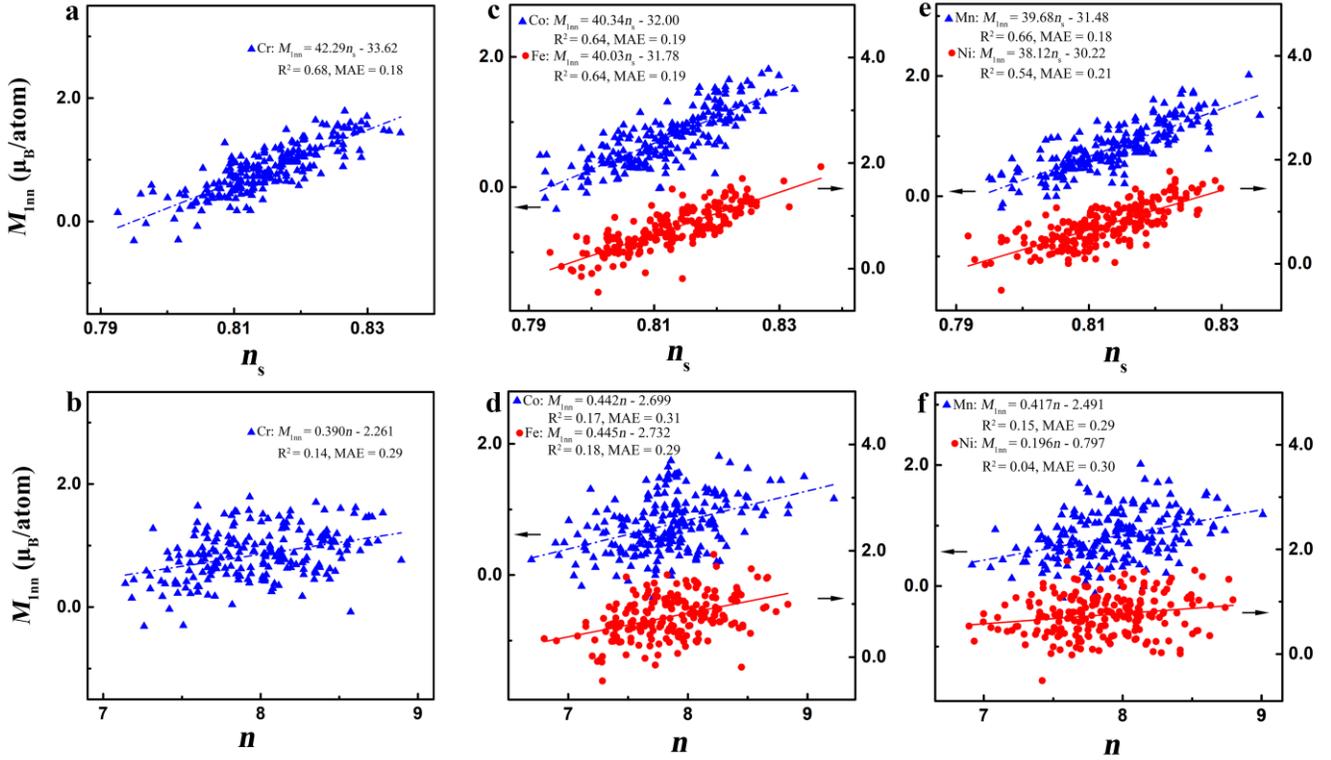

**Figure 4.** Comparison between the s-band filling ($n_s$) and total band filling ($n$) in describing the average magnetic moments of each vacancy's 1nn-shell atoms of CrMnFeCoNi HEAs. **a** and **b**, Cr vacancies. **c** and **d**, Co and Fe vacancies. **e** and **f**, Mn and Ni vacancies.

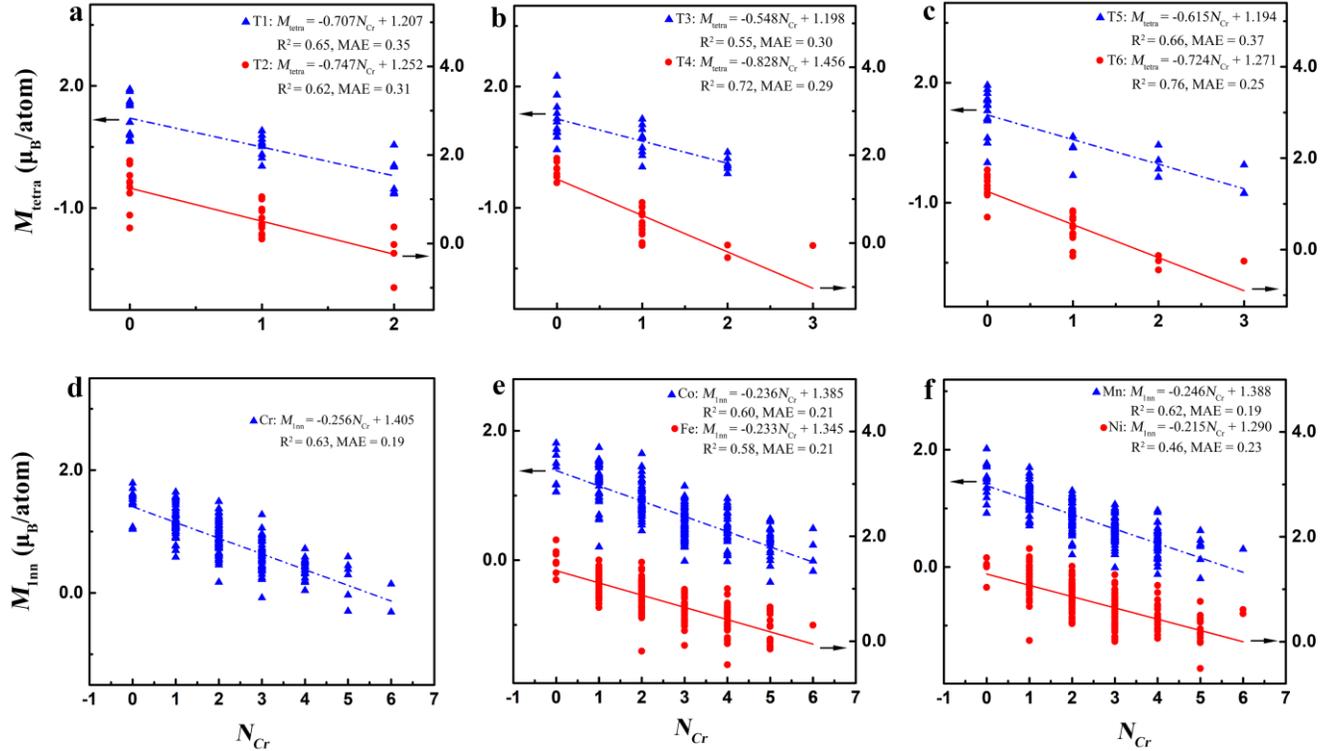

**Figure 5.** The average magnetic moments of CrMnFeCoNi HEAs as a function of the corresponding number of Cr atoms ($N_{Cr}$). **a-c**, Tetrahedrons of six different bulk structures (T1-T6). **d-f**, The 1nn-shell atoms of Cr, Co, Fe, Mn, and Ni vacancies.

and Cr atoms and between Cr and Mn atoms but not between other atoms in CrMnFeCoNi HEAs (Fig. 6). These localized electrons generate a covalent-like bonding for Cr-Cr pairs and Cr-Mn pairs, but not for the other atom pairs such as Cr-Fe, Cr-Co, and Cr-Ni pairs. Consequently, the mixed metallic and covalent-like bonding for Cr-Cr pairs and Cr-Mn pairs makes the d-state cohesive energy of Cr atoms in Cr-Cr pairs and Cr-Mn pairs increase from

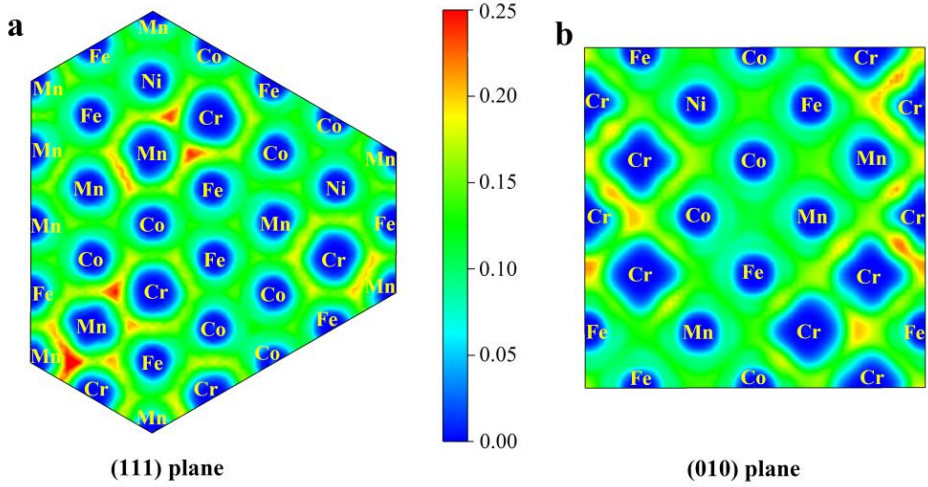

**Figure 6. The electron localization function (ELF) of CrMnFeCoNi HEAs projected on (111) (a) and (010) (b) planes.** The region with the deeper red color indicates the more electrons accumulated.

~ -8 eV to ~ -4 eV, which is just similar to the value of s-state cohesive energy $E_{coh-s}$ of bulk Cr atoms. Meanwhile, the other atom pairs experience a less change and their component atoms exhibit the d-state bonding highly similar to that in the bulk state. This explains well the proposed cohesive-energy descriptors for the chemical-environment effect of different sites in CrMnFeCoNi HEAs (Fig. 1). Overall, it is the localized electrons that change the d-states of Cr-Cr and Cr-Mn pairs and the d-states of atoms in CrMnFeCoNi HEAs still determine the chemical-environment effect, complying with the conventional TB and Friedel models[27,28]. Notably, for IrPdPtRhRu, NbMoTaTiV, and NbMoTaTiW HEAs, the d-state properties of component atoms in the bulk state always describe well the effect of chemical environments.

The local magnetic moments of CrMnFeCoNi HEAs can also be elucidated by the localized electrons between Cr and Cr atoms and between Cr and Mn atoms, which most likely stem from a part of itinerant electrons. This electron localization of Cr and Mn decreases the number of itinerant electrons and induces the energy level splitting and pronounced magnetic moments of Cr and Mn. In contrast, Fe, Co and Ni experience a less change of itinerant electrons as well as a minor change of magnetic moments. Note that Fe, Co and Ni atoms accept itinerant electrons from Cr and Mn atoms, and their electron-transfer amount reflects the change of their average magnetic moments (Supplementary Table 5). Moreover, in the 1nn shell of a given vacancy, the average magnetic moments of Cr atoms are related to those of the other atoms (Supplementary Fig. 11). Hence, the Cr atoms dominate the local magnetic moments of CrMnFeCoNi HEAs, as we found in Figs. 3-5. These results indicate that the localized and transferred itinerant electrons between different atoms control the magnetic properties of CrMnFeCoNi HEAs.

In summary, we have identified the mixed d- and s-state cohesive energy and s-band filling as effective descriptors in determining the trends of chemical bonding and magnetic moments of CrMnFeCoNi HEAs, based on the TB and Friedel models. The effectiveness of s-state properties is due to the unique role of Cr atoms and is absent in precious and refractory HEAs. The underlying mechanism is that the localized and transferred itinerant electrons in CrMnFeCoNi HEAs change the d-states of Cr atoms in Cr-Cr and Cr-Mn pairs and correlate the magnetic moments of Cr atoms with those of the other atoms. These results build a novel picture for the chemical bonding and magnetic moments of HEAs, which resolves the controversy between the chemical bonding and magnetic frustration of Cr-Cr pairs in CrMnFeCoNi HEAs, outlines the bonding characteristics of different HEAs, and supports the existence of SRO. Our descriptors can be alternative to the widely used concept, the chemical SRO parameters, by estimating the site-to-site bonding and magnetic moments, and serve as a design tool to engineering HEAs' properties.

**Methods**

In this study, we performed the DFT calculations via Vienna Ab initio Simulation Package (VASP) code[37] with the projector augmented wavefunctions (PAW) potentials[38] and Perdew−Burke−Ernzerhof (PBE) functional[39]. The quinary 3×3×3-supercell (containing 108 atoms) CrMnFeCoNi and IrPdPtRhRu bulk structures with near-equiatomic chemical compositions and approximately random distributions were built by using the Atomic Simulation Environment (ASE) and Python package PyCharm[40]. The lattice constant of a unit cell was 3.600 for CrMnFeCoNi HEAs and 3.911 for IrPdPtRhRu HEAs, which are consistent with the experimentally estimated results[41,42]. To study the chemical environments and magnetic moments of HEAs, both the non-spin-polarized and spin-polarized calculations were carried out. The $H_{vf}$ of each site was calculated to characterize the stability of HEAs, as it effectively connects the local site's stability with its surrounding chemical environments and magnetic moments. For the chemical-environmental effect, we calculated the non-spin-polarized $H_{vf}$ of each site of different CrMnFeCoNi and IrPdPtRhRu structures. Note that the results of NbMoTaTiV and NbMoTaTiW HEAs were cited from ref. [29]. For the magnetic effect, the magnetic moments of each atom of CrMnFeCoNi HEAs with and without vacancies were calculated, and the corresponding spin-polarized $H_{vf}$ was also obtained. The plane-wave cutoff energy of 500 eV and the $2 \times 2 \times 2$ Monkhorst-Pack k-point sampling mesh were adopted for geometry optimization. The valence electrons were specified as $3d^54s^1$, $3d^64s^1$, $3d^74s^1$, $3d^84s^1$, $3d^84s^2$ for Cr, Mn, Fe, Co, and Ni of CrMnFeCoNi HEAs, and $5d^86s^1$, $4d^95s^1$, $5d^96s^1$, $4d^85s^1$, $4d^75s^1$ for Ir, Pd, Pt, Rh, and Ru of IrPdPtRhRu HEAs. The conjugate gradient algorithm was utilized with a convergence threshold of 0.02 eV/Å in Hellmann-Feynman force on each atom. Note that we also employed a smearing width of 0.1 eV of the Fermi smearing function to facilitate the convergence. The electronic structures were analyzed by using the ELF[43] and Bader charge analysis[44].

The $H_{vf}$ was calculated as,

$$H_{vf} = E_{N-1} - E_N + \mu_{atom} \quad (3)$$

where $E_N$ and $E_{N-1}$ represent the total energies of bulk structures without vacancy and with one monovacancy respectively. $\mu_{atom}$ is the chemical potential of the given vacant atom of HEAs. Note that the chemical potential of an atom in HEAs highly depends on the interaction between the given atom and its surrounding environments, thus the determination of chemical potential of HEAs is challenging and prohibitively requires a high computational cost[29,45]. However, for a specific element, the variation trends of its $H_{vf}$ are independent of the value of chemical potential. We thus simply substituted $\mu_{atom}$ with the total energy of a vacant atom in its pure bulk phase to obtain $H_{vf}$ for characterizing the chemical-environment and magnetic effects of a given species in HEAs.

**Acknowledgments**
The authors are thankful for the support from the National Natural Science Foundation of China (Nos. 22173034, 11974128, 52130101), the Opening Project of State Key Laboratory of High Performance Ceramics and Superfine Microstructure (SKL201910SIC), the Program of Innovative Research Team (in Science and Technology) in University of Jilin Province, the Program for JLU (Jilin University) Science and Technology Innovative Research Team (No. 2017TD-09), the Fundamental Research Funds for the Central Universities, and the computing resources of the High Performance Computing Center of Jilin University, China.

**Author contributions**
W. G. and Q. J. conceived the original idea and designed the strategy. B. L. performed the DFT calculations. W. G. derived the models and analyzed the data with the contribution from B. L.. B. L., W. G. and Q. J. discussed the results and wrote the manuscript together. B. L. prepared the Supplementary Information and drew all figures. All authors have approved the final version of the manuscript.

**Competing interests**
The authors declare no competing interests.